\documentclass[11pt]{article}

\usepackage{upgreek}
\usepackage{siunitx}

\usepackage[T1]{fontenc}
\usepackage[utf8]{inputenc}
\usepackage{textalpha}
\usepackage[scale=0.92]{roboto}
\usepackage[charter]{mathdesign}
\DeclareFontFamilySubstitution{LGR}{\rmdefault}{txr}

\usepackage[a4paper,margin=1.0in]{geometry}
\usepackage{setspace}
\usepackage{titlesec}
\titleformat*{\section}{\Large\bfseries\sffamily}
\titleformat*{\subsection}{\large\bfseries\sffamily}
\titleformat*{\subsubsection}{\normalsize\bfseries}

\usepackage{authblk}
\usepackage{amsmath}
\usepackage{graphicx}
\usepackage{multirow, booktabs}
\usepackage[hidelinks]{hyperref}
\usepackage{xurl}
\usepackage[articletitle=true]{achemso}
\usepackage{caption}
\usepackage{siunitx}

\usepackage[defaultcolor=red]{changes} 

\newcounter{suppfigure}
\newenvironment{suppfigure}[1][]
{
    \refstepcounter{suppfigure}
    \begin{figure}[#1]
    
}
{
    \end{figure}
}

\title{
\raggedright
	\centering
	\sffamily \Large \textbf{GPx4 is bound to peroxidized membranes by a hydrophobic anchor}}

\author[a,1]{Qingyang Hu}
\author[b,1]{Hantian You}
\author[b]{Kenan Li}
\author[a,b,c,*]{Luhua Lai}
\author[a,c,*]{Chen Song}

\affil[a]{Center for Quantitative Biology, Academy for Advanced Interdisciplinary Studies, Peking University, Beijing 100871, China}
\affil[b]{Beijing National Laboratory for Molecular Sciences, College of Chemistry and Molecular Engineering, Peking University, Beijing 100871, China}
\affil[c]{Peking-Tsinghua Center for Life Sciences, Academy for Advanced Interdisciplinary Studies, Peking University, Beijing 100871, China}

\affil[1]{These authors contributed equally to this work.}
\affil[*]{E-mail: \href{mailto:lhlai@pku.edu.cn}{lhlai@pku.edu.cn} (L.L.), \href{mailto:c.song@pku.edu.cn}{c.song@pku.edu.cn} (C.S.)}

\date{}

\begin{document}

\doublespacing

\maketitle

\clearpage
\onehalfspacing




\begin{abstract}
Ferroptosis is a form of cell death discovered in recent years, induced by excessive peroxidation of phospholipids. Glutathione peroxidase 4 (GPx4) is an intracellular enzyme that can repair the peroxidized phospholipids on membranes, thus regulating ferroptosis. By combining multiscale molecular dynamics (MD) simulations and experimental assays, we investigate the binding mechanisms of GPx4 on membranes. Using coarse-grained MD simulations, we found that L130 and its adjacent residues on GPx4 can form a stable and unique binding interface with PE/PS-rich and peroxidized membranes. Subsequent all-atom MD simulations verified the stability of the binding interface. The critical residue on the interface, L130, was inserted deeply into the membrane as a hydrophobic anchor and guided the reaction center toward the membrane surface. Enzyme activity assays and \textit{in vitro} cell experiments showed that mutations of L130 resulted in weaker activities of the enzyme, probably caused by non-functional binding modes of GPx4 on membranes, as revealed by \textit{in silico} simulations. This study highlights the crucial role of the hydrophobic residue, L130, in the proper anchoring of GPx4 on membranes, the first step of its membrane-repairing function.
\end{abstract}












\section{Main}

Cell death is an essential process during the growth and development of living organisms. The regulation of cell death is crucial for metabolism, immunoregulation and homeostasis. Ferroptosis, coined in 2012, is an iron-dependent and regulated cell death modality induced by phospholipid peroxidation \cite{2012_Ferroptosis}. It is distinctive from other cell death modalities, and many diseases such as diffuse large B cell lymphoma (DLBCL),  Huntington's disease, and Periventricular Leukomalacia (PVL), are related to its misregulation \cite{2016_FerroptosisReview, 2021_FerroptosisDiseases}. Using inducers targeting ferroptosis to treat cancers has been in full swing recently because plenty of iron is demanded in the rapid growth of cancer cells, making phospholipids under great peroxidative pressure \cite{2022_TreatCancer}.

Several molecular pathways for ferroptosis regulation have been discovered in recent years \cite{2021_FerroptosisPaths}, among which glutathione peroxidase 4 (GPx4) plays a crucial role in the most classic pathway since it reduces phospholipid peroxides to alcohol by oxidizing glutathiones (GSHs) to glutathione disulfides (GSSGs) \cite{1997_GPx4withGSH} (Fig. 1A). Compared with other glutathione peroxidases (GPx) containing selenocysteine (U), GPx4 is the only known one that can treat phospholipid peroxides as redox reaction substrates \cite{2003_onlyGPx4}. Therefore, GPx4 is an essential enzyme that can inhibit ferroptosis by both preventing and repairing cell membranes from the accumulation of peroxides.

To understand the ferroptosis process, computational studies were conducted to study the destructive effect of phospholipid peroxides on cell membranes \cite{2016_MD_Phospholipids-OOH, 2016_MD_Phospholipids-OOH_Membrane, 2018_LipidPeroxidation}. Docking calculations showed the likely interaction mode between the GPx4 and lipid bilayers \cite{2017_GPx4}, which was in line with the X-ray structure-based analysis, where three key residues, U46, Q81, and W136, form a catalytic triad that should be located at the protein surface toward membranes \cite{2007_GPx4-U46C}. Recently, electrostatics was found to be crucial for driving GPx4 interactions with membranes \cite{2021_GPx4-NMR} and the R152H mutant could disturb the interaction of the enzyme with membranes \cite{2021_GPx4-R152H,2023_GPx4-R152H}. However, a high-resolution picture of GPx4 interacting with membranes is still needed to fully reveal the dynamic anchoring process of GPx4 on peroxidized membranes, which would help develop new regulation strategies.

In this work, by using multiscale molecular dynamics (MD) simulations (Table S1) and experimental assays, we revealed the detailed physical interactions between the enzymatically active U46C mutant \cite{2000_U46C} of GPx4 and membranes composed of POPC, POPE, POPS, and their peroxidized derivatives (termed HPPC, HPPE and HPPS) (Fig. 1B). Coarse-grained (CG) MD simulations (Fig. S1A--B) were used to study the dynamic binding process of GPx4 on membranes composed of different lipid types (Fig. 1C), which showed a stable and unique binding configuration on PE/PS-rich membranes. Based on the stable binding configuration, all-atom (AA) MD simulations (Fig. S1C--D) were conducted to further evaluate the stability of the binding interface, which also validated L130 was the most important membrane-inserted residue on the GPx4 surface. L130 appeared to serve as a hydrophobic anchor that penetrated deepest into the hydrophobic core of the membrane and stabilized the reaction center of GPx4 toward the membrane surface. Enzyme activity assays and \textit{in vitro} cell experiments provided further validations, highlighting the important role of the hydrophobic residue L130 in stabilizing the anchoring configuration of GPx4 on membranes, which was also proved by \textit{in silico} simulations.

\begin{figure}[htbp]
    \centering\includegraphics[width=7cm]{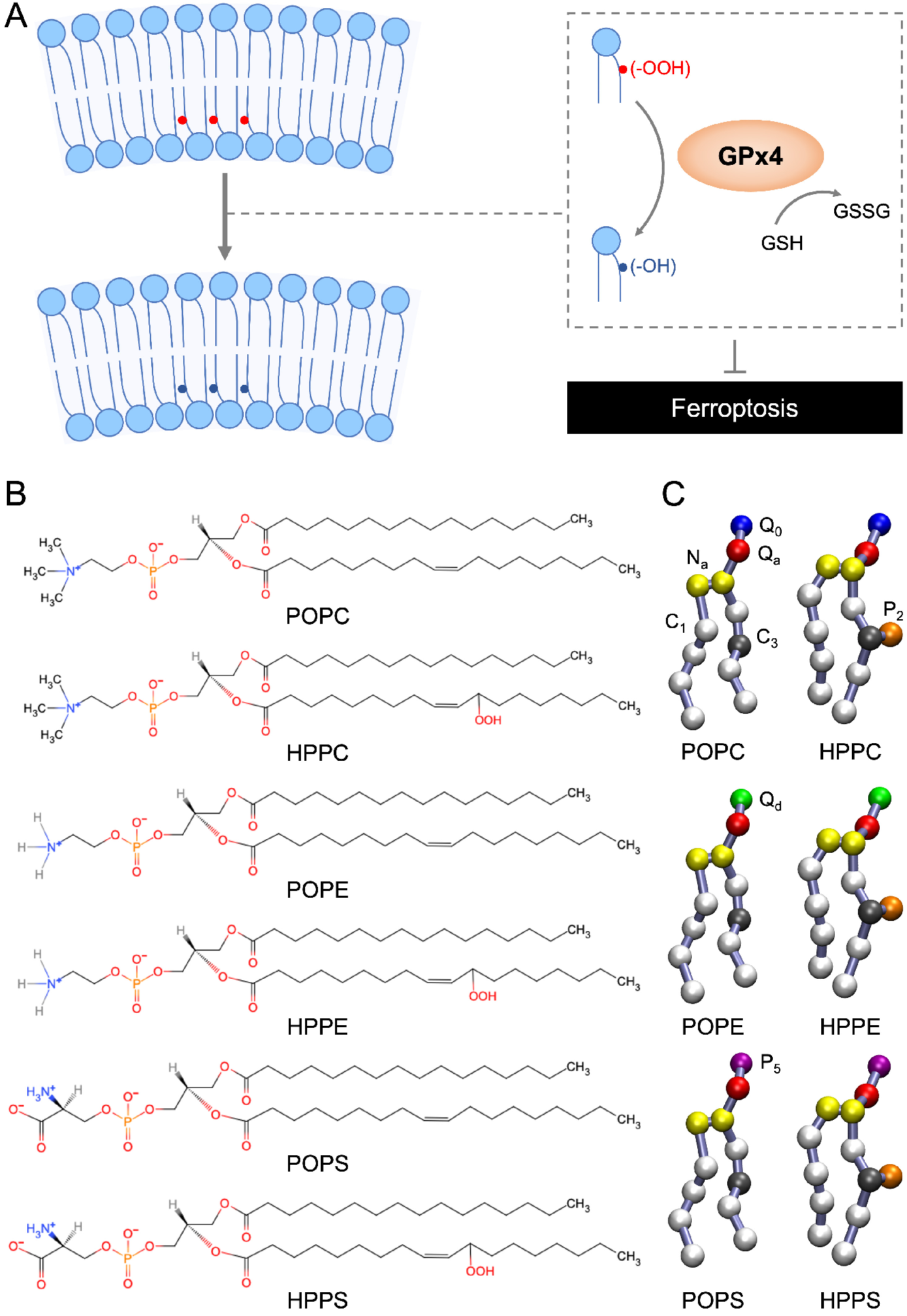}
    \caption{(A) The key role of GPx4 in the ferroptosis regulation. By oxidizing GSHs to GSSGs, GPx4 can reduce phospholipid peroxides of the inner leaflet to alcohol, thus inhibiting ferroptosis. The unsaturation of membrane phospholipids is asymmetric, as the inner leaflet is approximately twofold more unsaturated than the external leaflet \cite{Lorent2020}. (B) AA models and (C) CG models of the simulated phospholipids, i.e. POPC, POPE, POPS, and their corresponding peroxides HPPC, HPPE, and HPPS, respectively. GSH, glutathione; GSSG, glutathione disulfide; GPx4, glutathione peroxidase 4; POPC, 1-palmitoyl-2-oleoyl-\textit{sn}-glycero-3-phosphocholine; HPPC, peroxidized POPC; POPE, 1-palmitoyl-2-oleoyl-\textit{sn}-glycero-3-phosphoethanolamine; HPPE, peroxidized POPE; POPS, 1-palmitoyl-2-oleoyl-\textit{sn}-glycero-3-phospho-L-serine; HPPS, peroxidized POPS; Q, charged CG bead; N, nonpolar CG bead; C, apolar CG bead; P, polar CG bead; 0, bead of no hydrogen-bonding capabilities; a, bead acting as hydrogen bond acceptor; d, bead acting as hydrogen bond donor; 1--5, indicating the polar affinity of bead, whose maximum value is 5.}
    \label{fig:GPx4&lipids}
\end{figure}

Our CG MD simulations showed that there was a stable binding mode of GPx4 on membranes, and the binding affinity depends on the compositions of membranes. As can be seen in Fig. 2A--D, when the lipid bilayer was composed of pure POPC, there was no stable binding interaction between GPx4 and the bilayer, since the distance between the centers of the mass (COMs) of the two was fluctuating between 4 nm and 7 nm all the time (Fig. 2A), where a distance of less than 4 nm indicates the GPx4 being attached to the membrane surface. In contrast, GPx4 could stably bind to the surface of the POPC/POPE membrane and the POPC/POPE/POPS membrane with COMs distances of $\sim$3.5 nm (Fig. 2B--C). Furthermore, we converted the head groups of POPE and POPS to POPC based on the stable binding pose of GPx4 on the POPC/POPE/POPS membrane after the 50-μs CG simulation and conducted three extended 10-μs CG simulations. As can be seen, the distance between the COM of GPx4 and the membrane midplane is always fluctuating and mostly greater than 4 nm after 1 μs (Fig. S2), indicating that GPx4 was detached from the mutated membrane. Therefore, the interaction between GPx4 and the pure POPC membrane is indeed unstable in the simulations.

The difference in the binding affinity may arise from the different types of the head groups of these phospholipids, representing their different features in physical chemistry. Although the head groups of PC/PE/PS are all represented by one bead in the Martini force field, they have different non-bonded interaction parameters, with POPC being less attractive or more repulsive than POPE and POPS \cite{2007_MARTINI}. This may correspond to the fact that the PC head group is larger than PE or PS in the atomic model, so that the head group of POPC can lead to a more compact hydrophilic wall and less space for GPx4 to be inserted and reach the hydrophobic core of the lipid bilayer. Notably, the binding of GPx4 to the membrane with peroxidized phospholipids showed high affinity as well, where the GPx4 was never detached from the membrane surface once it was stably bound in our simulation (Fig. 2D). 

\begin{figure}[htbp]
    \centering
    \includegraphics[width=12.5cm]{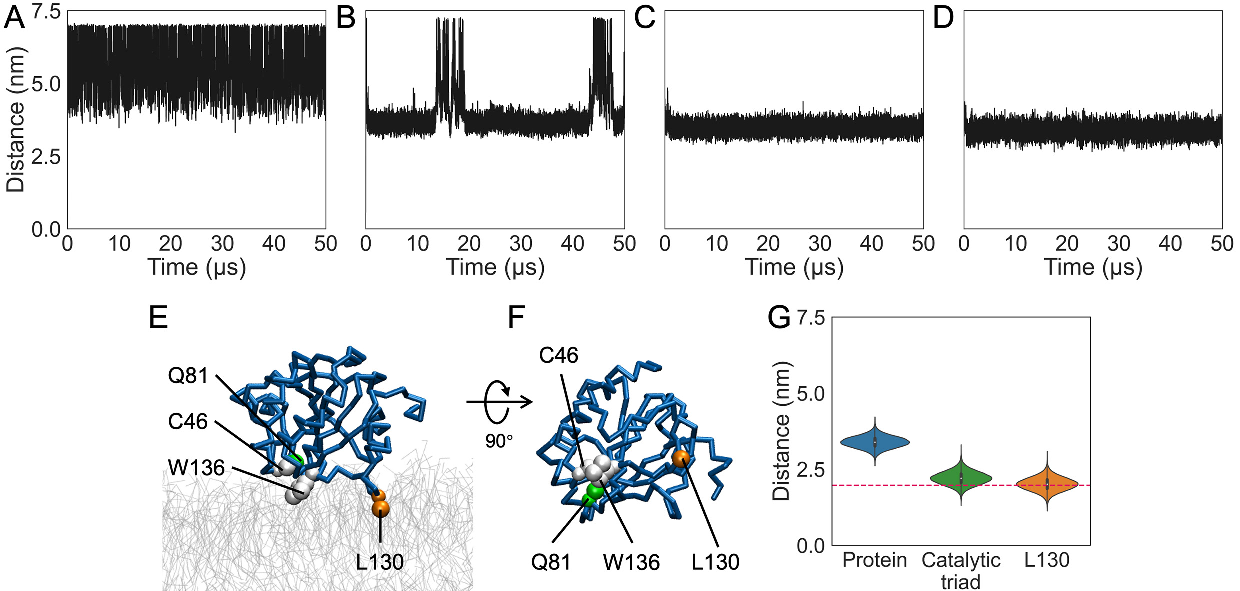}
    \caption{Binding stability of GPx4 with different membranes. (A-D) Distances between GPx4 and various membranes in coarse-grained MD simulations: (A) the pure POPC membrane, (B) the POPC/POPE membrane (POPC:POPE=1:1), (C) the POPC/POPE/POPS membrane (POPC:POPE:POPS = 1:2:2), and (D) the mixed membrane containing peroxidized phospholipids (POPC:POPE:POPS:HPPC:HPPE:HPPS = 1:2:2:1:2:2). (E--F) The interaction interface of GPx4 with the membrane containing peroxidized phospholipids after a 50-$\upmu$s CG MD simulation. (G) Distributions of the distances between the protein, the catalytic triad, and L130 and the mixed membrane containing peroxidized phospholipids. The distances in (A-D) were calculated between the center of mass of the protein or the residue and the membrane midplane. The data in (G) were analyzed over the 10–50 $\mu$s range of (D) and the dotted line represented the average location of the head groups of lipids ($\sim$1.97 nm).}
    \label{fig:affinity}
    
\end{figure}

The interaction interface of GPx4 with the peroxidized membrane did not change in our 50-$\upmu$s simulations, with the catalytic triad facing toward the membrane (Fig. 2E--G), indicating that the binding was stable and the interface was perhaps specific. To verify this, we approximated the protein as a cube and chose its six different initial orientations on membranes, or rather, binding interfaces, and performed six 10-$\upmu$s CG MD simulations for each case. The simulation results showed that GPx4 kept rolling on the membrane initially and sometimes detached from the membrane, until most trajectories converged to the same stable binding interface, the sixth initial configuration ([6] in Fig. 3A), a state with the catalytic triad in direct contact with the membrane surface (Figs. 3 and S3). This binding mode corresponded to the configuration observed in the aformentioned 50-$\upmu$s CG simulations. Analysis of six independent trajectories originating from this configuration revealed that L130 was the most deeply membrane-embedded residue within the lipid bilayer throughout all simulations (Fig. S4). Apparently, L130 served as a hydrophobic anchor embedded deeply within the hydrophobic core of the lipid bilayer, which was crucial for stabilizing the binding of GPx4 to the membrane. Meanwhile, the reaction center residues, U46/C46, Q81, and W136 were oriented toward the membrane surface in this configuration, thereby facilitating their interaction with peroxidized lipids to repair the damaged membrane.

\begin{figure}[htbp]
    \centering
    \includegraphics[width=12.5cm]{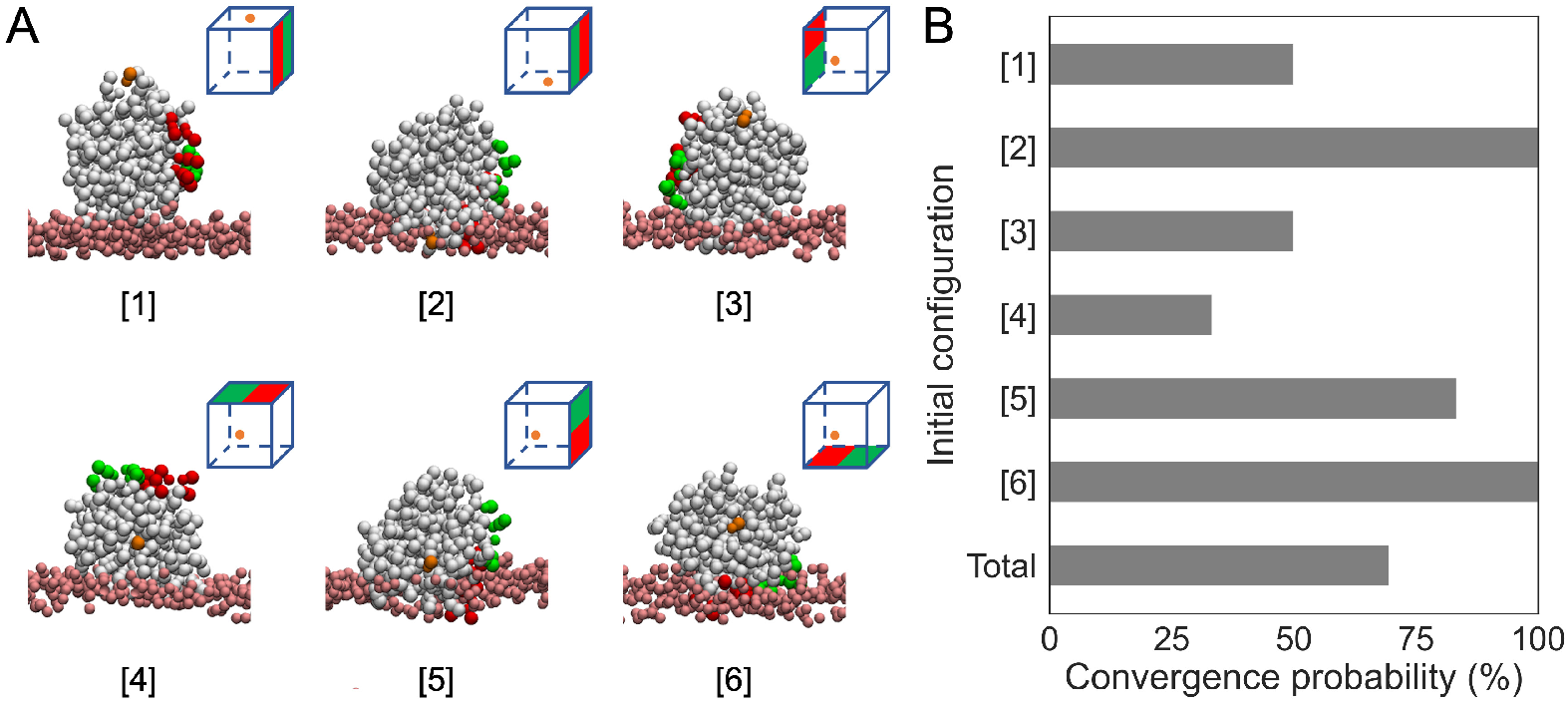}
    \caption{The most probable binding interface in the CG simulations with various starting configurations. (A) Initial configurations visualized with a cube analogue. (B) Convergence probability from different initial configurations to the sixth configuration that is most stable ([6] in panel A). Six trajectories are counted for each initial configuration, and 36 trajectories are counted in total. The red residues are I129-N132 and K135-W136, the green ones are F78-K80 and P83, and the orange one is C10. These residues are used to indicate different orientations of the protein on the membrane surface. All simulations here used mixed membranes containing peroxidized phospholipids, with the POPC:POPE:POPS:HPPC:HPPE:HPPS composition ratio being 1:2:2:1:2:2.}
    \label{fig:binding_interface}
\end{figure}

With the stable binding interface identified in CG MD simulations, we further performed atomistic MD simulations to reveal the details of the GPx4-membrane interactions.
Here, three different AA systems were built for simulations. Firstly, the final configuration of one of the 10-$\mu$s CG simulations from the sixth initial configuration (Fig. S3F) was converted into an AA model, in which L130 inserted deeply into the interior of the membrane (Fig. S5A). Secondly, a CG configuration, in which L130 had not yet inserted into the membrane but was very close to the membrane surface, was converted into an AA model (Fig. S5B). The third system is a duplication of the first system except with a L130S mutation (Fig. S5C).

In the first system, atomistic simulation results validated that GPx4 was indeed stably bound to the membrane surface and the interaction surface showed a high degree of similarity to the CG results (Fig. S5D). Specifically, L130 was the residue that inserted most deeply into the membrane and the catalytic triad was exposed to lipid molecules (Fig. 4A--B).
In the second system, GPx4 eventually formed the stable binding with the peroxidized membrane in two out of the three trajectories (Fig. S5E). However, in one of these trajectories, L130 failed to insert into the membrane, causing GPx4 to be slightly separated from the membrane. Such results indicate that if L130 fails to insert into the interior of the peroxidized membrane, GPx4 will eventually separate from the peroxidized membrane. 
In the third system with the L130S mutation, GPx4 was not observed to dissociate from the membrane in the accessible simulation time (Fig. S5F). However, it can be seen that the insertion depth of the protein, the catalytic triad and S130 were shallower than in the first atomistic system, indicating a less stable binding.
Taken together, these results indicated that L130 was the most critical residue for the stable anchoring of GPx4 on membranes, as it can reach the hydrophobic core of membranes, facilitating contact between the residues around it (including the catalytic triad) and the membrane surface (Fig. S6).

\begin{figure}[htbp]
    \centering
    \includegraphics[width=12.5cm]{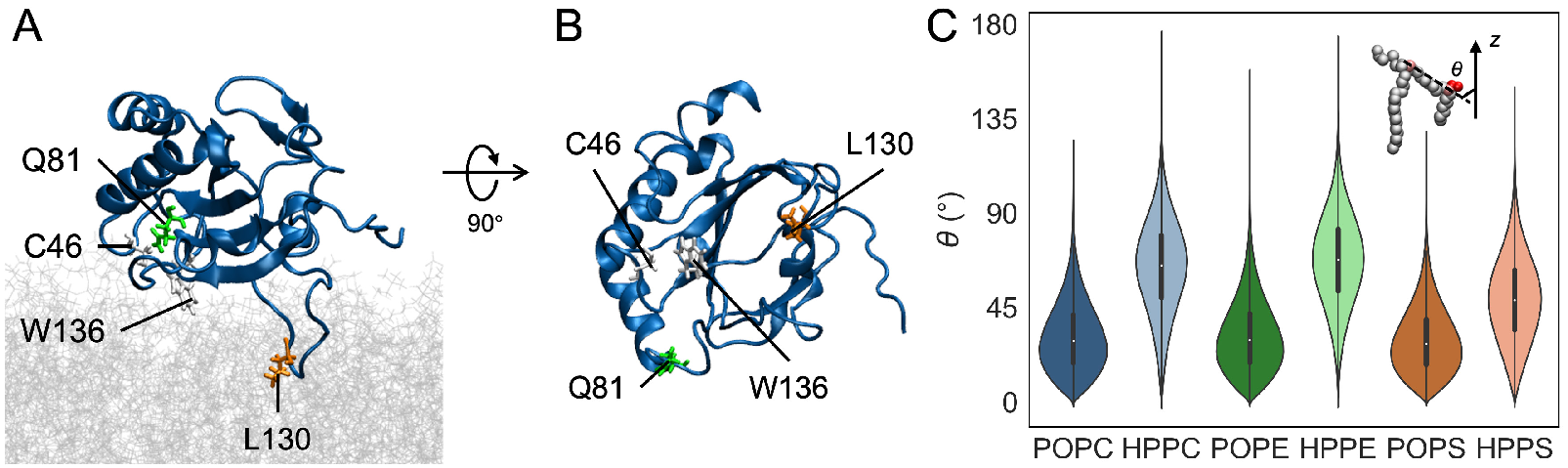}
    \caption{The GPx4-membrane interaction in all-atom simulations. (A-B) The interaction interface of GPx4 with the membrane containing peroxidized phospholipids after a 500-ns AA MD simulation. (C) Distributions of deviation angles of different phospholipids. The deviation angle is defined in the inset, where the dotted line extends from the carbon atom attached to the head group (pink bead) to the carbon atom attached to the peroxide group (pink bead attached to the red bead) and the $z$-axis represents the membrane normal direction.}
    \label{fig:conf_hppe}
\end{figure}

In addition, our simulations showed that peroxidized lipids have more tilted tails than normal lipid molecules. As shown in Fig. 4C, the deviation angle, defined as the angle between the $z$-axis (membrane normal direction) and the line through the carbon atom attached to the head group and the carbon atom attached to the peroxide group. The average values of the angles within the simulation time were approximately \qty{31}{\degree}, \qty{32}{\degree} and \qty{30}{\degree} for POPCs, POPEs, and POPSes, respectively. In contrast, the tails of peroxidized lipids (HPPCs, HPPEs, and HPPSes) tend to tilt and deviate further from the membrane's normal direction, resulting in a larger deviation angle of around \qty{64}{\degree}, \qty{68}{\degree}, and \qty{49}{\degree}, respectively. Analyses of lipid order parameters and carbon atom densities confirmed that the peroxidized lipids exhibit a more disorded structure than the normal ones (Fig. S7). These were consistent with previous studies showing that the bending of the peroxidized tail was an inherent property of peroxidized phospholipids \cite{2016_MD_Phospholipids-OOH, 2019_POPCOOH}. Thus, the peroxidized lipid molecules, when lifted in such a way, become more exposed on the membrane surface and are available for interaction with membrane-bound GPx4. Therefore, from the above simulation results, it is reasonable to speculate that L130 can anchor GPx4 on the damaged membrane surface so that its catalytic residues can better capture the 'lifted' peroxidized group of lipid molecules to execute the repairing function.

As GPx4 can inhibit ferroptosis by binding and repairing peroxidized cell membranes, we utilized peroxidized liposomes to assess the activity of GPx4 and to evaluate whether mutations at the hydrophobic L130 residue diminish GPx4's activity to repair membranes. Liposomes were prepared with a phospholipid mixture and enclosed 5-Carboxyfluorescein \cite{1984_Liposome}, followed by incubation with purified ALOX15, which was sufficient to cause leakage from the liposomes. The fluorescein within the liposomes exhibits a weak signal and the signal intensity will increase significantly if it is released from the liposomes. Here, we also used the enzymatically active GPx4-U46C mutant \cite{2000_U46C} to substitute for the wild type of GPx4. GPx4 addition suppressed ALOX15-induced liposome leakage, confirming its membrane repair activity (Fig. S8). Substitution of the hydrophobic L130 residue with polar residues (e.g., L130S and L130Q) markedly reduced GPx4's membrane repair capability (Fig. 5A, left panel). Importantly, these mutations did not decrease the activity of GPx4 in simple soluble substrates in solution (Fig. 5A, right panel). Given that the expression levels of GPx4 and its mutants were comparable (Fig. S9A), these results indicated that the weakening of membrane repair ability stems from impaired membrane binding of the mutated GPx4.

Furthermore, we conducted cell experiments to explore whether the mutations of L130 can lead to the failure to suppress ferroptosis induced by erastin. Quantification of cell viability using cell proliferation assay revealed a considerable increase in the sensitivity of GPx4 knockout 786-O cells to erastin, which was rescued by expression of the wild type (U46C) and its L130A mutant. However, the L130S and L130Q mutations significantly impair the ability of GPx4 to inhibit ferroptosis. Therefore, our experimental results confirmed that hydrophobic L130 is essential for the proper function of GPx4 (Figs. 5B, S9B, and S10).

To investigate the underlying mechanisms responsible for the diminished function of the above mutants, we conducted corresponding \textit{in silico} mutation simulations. CG methods were employed to investigate the dynamic interactions of GPx4-U46C and its mutants (U46C/L130A, U46C/L130S, and U46C/L130Q) with the peroxidized membrane (POPC: POPE: POPS: HPPC: HPPE: HPPS = 1:2:2:1:2:2). Initially, the protein was randomly placed in a position close to the membrane surface (Fig. S11A). For different mutants, the initial positions of all atoms were the same, except that the residue L130 was mutated into A130, S130, and Q130, respectively. Three 10-$\upmu$s CG MD simulations were performed for each case. The distances between different mutants, their catalytic triads, and the 130th residues and the membrane midplane were calculated and shown in Fig. S11B--E.

The CG MD results indicated that, within our simulation time, only U46C and U46C/L130A were able to deeply bind to the peroxidized membrane and form an effective binding interface, with L130 inserting into the interior of the membrane (Fig. 5C, left panel) and the catalytic triad approaching the membrane surface (Fig. 5C, middle panel). For mutants U46C/L130S and U46C/L130Q, although they did not dissociate from the membrane (Fig. 5C, right panel), their catalic triads were not as close to the membrane surface as U46C. Therefore, we speculate that U46C and U46C/L130A are able to effectively and stably bind to the peroxidized membrane, guiding the catalytic triad of GPx4 to approach the membrane surface and contact the peroxidized lipids, while mutants such as U46C/L130S and U46C/L130Q are less efficient at doing so. These can well explain our \textit{in vitro} experimental results.

\begin{figure}[htbp]
    \centering
    \includegraphics[width=12.5cm]{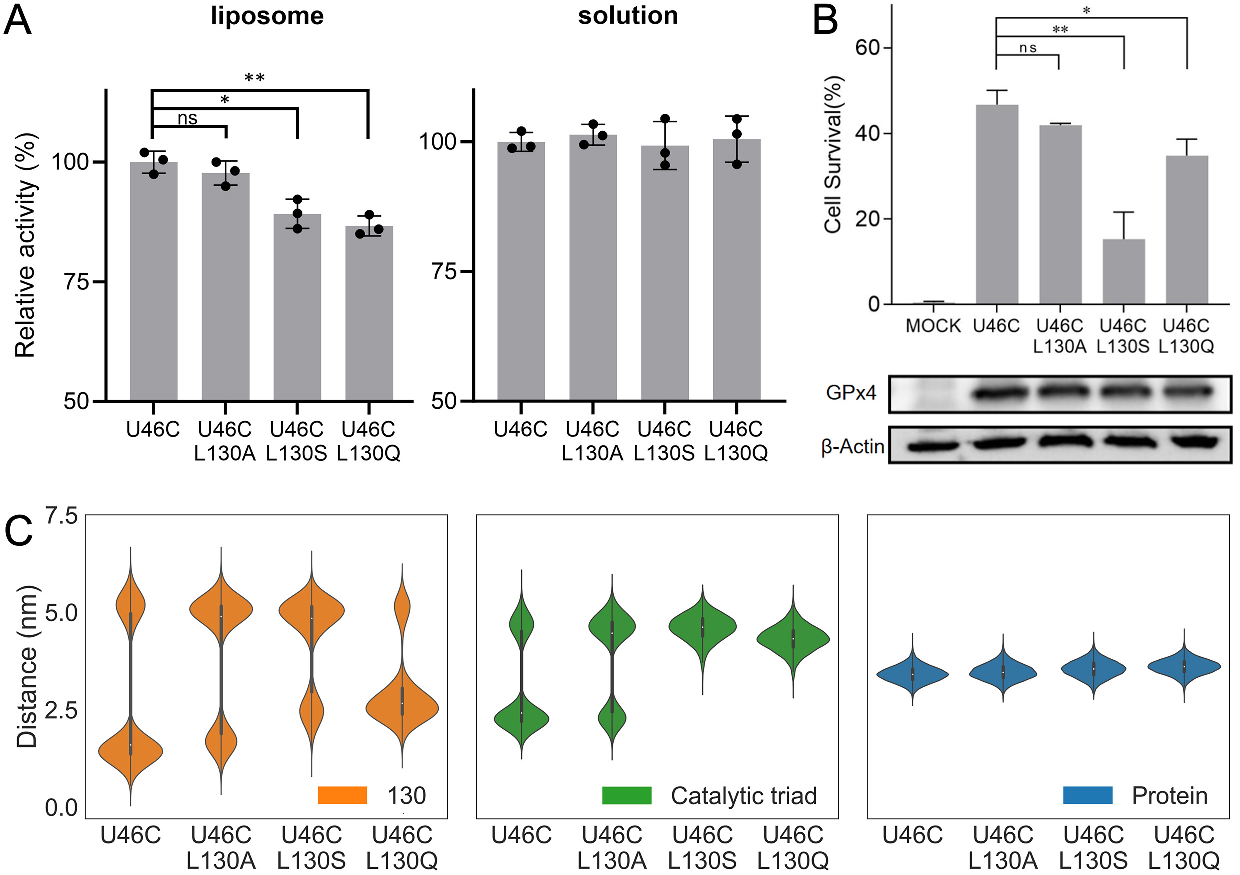}
    \caption{Validation of the key role of L130 on the binding interface. (A) Activity of GPx4 and its mutants to liposomes and soluble substrates; *P<0.05, **P<0.01, NS, not significant. (B) Cell viability in GPx4-knockout 786-O cells overexpressing mock, GPx4(U46C), GPx4(U46C/L130A), GPx4(U46C/L130S) or GPx4(U46C/L130Q) treated with 2.5 $\upmu$M erastin (top) and western blot analysis of GPx4 KO cells that express GPx4-U46C and mutations (bottom). (C) Distributions of distances between the 130th residues, the catalytic triads, and the protein centroids of GPx4 mutants and peroxidized membranes, resulting from \textit{in silico} simulations. The proteins were randomly placed in a position close to the membrane surface and the distances were calculated between the COMs of the protein atoms and the membrane midplanes. The analyses were conducted using the trajectories after 5 $\upmu$s.}
    \label{fig:validation}
\end{figure}

In summary, our results showed that: 1) GPx4 has different binding affinities with membranes of different compositions, showing a stronger binding affinity with PE and PS lipids than PC. It appears that GPx4 has stronger binding interactions with those small-head, less compact, and peroxidized phospholipids, probably owing to the steric effects. This makes GPx4 readily bind to the 'damaged' regions of membranes to execute its lipid-repairing function. 2) GPx4 has a unique binding interface with membranes, which exposes the reaction center to phospholipids. L130 appears to be a critical hydrophobic anchor that can stabilize the proper GPx4-membrane interaction interface, and it ensures that GPx4 can not only touch the membrane surface but also be inserted into the leaflet to reach the hydrophobic core of the membrane, thus more stably bound. In the meantime, the reaction center can reach the lifted peroxidized group of lipid molecules, facilitating the following reduction reactions.

Evidently, L130 is the most important residue for the proper binding of GPx4 on the damaged membranes. Mutations of L130 to polar residues, such as L130S or L130Q, lead to a weaker membrane-repairing function. Accordingly, disturbing the binding interface around L130 can be a new approach to suppress GPx4 and promote ferroptosis. 

\section{Materials and Methods}

\subsection{System setup}
The structure of GPx4 was downloaded from the Protein Data Bank (PDB ID: 2OBI), and then the CG model was generated by \emph{martinize.py} \cite{2013_martinize}. It should be noted that the structure of GPx4-U46C instead of GPx4 was used in our MD simulations due to the lack of parameters for Se in the available force fields, but GPx4-U46C is still enzymatically active \cite{2000_U46C} and the results were not expected to be significantly affected for our purpose because this would not alter the likely interaction interface between the GPx4 and membranes. In the virtual mutation validation, GPx4 U46C/L130A, U46C/L130S and U46C/L130Q were constructed with Chimera \cite{2004_Chimera}, and the rotamer library Dunbrack 2010 was used for modeling \cite{2011_Dunbrack}.

The model membranes used in our simulations were composed of 2-oleoyl-1-palmitoyl-sn-glycero-3-phosphocholine (POPC), 1-palmitoyl-2-oleoyl-sn-glycero-3-phosphoethanolamine (POPE), 1-palmitoyl-2-oleoyl-sn-glycero-3-phospho-L-serine (POPS) and their corresponding peroxides, named HPPC, HPPE and HPPS respectively (Fig. 1B). Parameter files of POPC, POPE, and POPS were readily available, while HPPC, HPPE, and HPPS were new types of phospholipids that needed to be parameterized. Fabrice Thalmann et al. developed the CG model and parameter file of HPPC (i.e. HP-POPC) \cite{2016_HPPCmodel} within the Martini2.2 force field \cite{2007_MARTINI}, and the parameters for HPPE and HPPS were created in the same way. The AA parameter files of HPPC, HPPE, and HPPS were first generated by the module \emph{Ligand Reader \& Modeler} \cite{2017_LigandModeler} in CHARMM-GUI \cite{2017_CHARMM36m}, and then the atom types within these files were further checked and adjusted to ensure that the parameters of HPPC, HPPE, and HPPS were consistent with those of POPC, POPE, and POPS, except for the additional peroxidized group and the connected carbon atom. Four types of phospholipid bilayers were constructed for our simulations: a pure POPC bilayer, a mixed bilayer consisting of POPC and POPE (POPC: POPE = 1:1), a mixed bilayer consisting of POPC, POPE, and POPS (POPC: POPE: POPS = 1:2:2), and a mixed POPC/POPE/POPS bilayer containing peroxidized phospholipids (POPC: POPE: POPS: HPPC: HPPE: HPPS = 1:2:2:1:2:2). All the simulated bilayers are symmetric, that is, for the same membrane, the phospholipid compositions of the inner and outer leaflets are the same. The phospholipid composition ratio of the POPC/POPE/POPS membrane is set close to the inner bilayer of plasma membranes \cite{Verkleij1973, VanderSchaft1987, Klähn2013, Ingólfsson2014, Lorent2020, Hazrati2025}.
The CG simulation systems were generated with \emph{insane.py} \cite{2015_insane} and solved in a water box of 12 nm \texttimes\ 12 nm \texttimes\ 15 nm (Fig. S1A). The script \emph{backward.py} \cite{2014_backward} was used to build AA models (Fig. S1C) from the last frame of the CG simulations.

\subsection{Molecular dynamics simulations}
CG simulations were performed with the GROMACS 2021.2 package \cite{2015_GROMACS} using the Martini2.2 force field with an elastic network \cite{2007_MARTINI}. To achieve energy minimization, the steepest descent algorithm was used for thousands of steps and the resulting maximum force on each atom was less than 10 kJ mol$^{-1}$ nm$^{-2}$. Then, the systems underwent a 500-ps equilibration to reach 310 K with position restraints on the backbone beads (force constant 1000 kJ mol$^{-1}$ nm$^{-2}$) in the NVT ensemble. Then the production simulations were performed under the NPT ensemble and the temperature and the pressure were maintained at 310 K and 1.0 bar by using the V-rescale algorithm \cite{2008_Gromacs4} with a time constant of 1.0 ps and the Berendsen algorithm \cite{1984_Berendsen} with a time constant of 10.0 ps, respectively. The Berendsen barostat was used to adapt to the possible changes in box size caused by the insertion of GPx4 into the membrane. In all the simulations, the periodic boundary conditions were applied and Newton's equations of motion were integrated at intervals of 10 fs. Electrostatics were calculated with a cutoff of 1.1 nm and a relative dielectric constant of 15, while the permittivity was set to infinity beyond the cutoff.

AA simulations were performed with the GROMACS 2021.2 package \cite{2015_GROMACS} and conducted with the CHARMM36m force field \cite{2017_CHARMM36m}. 0.15 mol/L NaCl was added to the AA system.
Before AA production simulations, an energy minimization with thousands of steps, a 500-ps NVT equilibration, and a 50-ns semi-isotropic NPT equilibration were conducted to reach a temperature of 310 K and a pressure of 1.0 bar, and position restraints were applied on C$\upalpha$ atoms of GPx4, with a force constant of 1000 kJ mol$^{-1}$ nm$^{-2}$. The production simulations were under the semi-isotropic NPT ensemble without any restraints, in which the V-rescale algorithm \cite{2007_v-rescale} with a time constant of 0.5 ps and the Parrinello-Rahman algorithm \cite{1981_Parrinello-Rahman} with a time constant of 1.0 ps were used to maintain the temperature at 310 K and the pressure at 1.0 bar, respectively. In all of AA simulations, the periodic boundary conditions were used and the time step was set to 2 fs. The short-range cut-offs were 1.2 nm for electrostatic and van der Waals (vdW) interactions, with a switch function turning off the vdW interactions from 1.0 to 1.2 nm. The long-range electrostatic interactions were calculated with the particle mesh Ewald (PME) method \cite{1993_PME}. 

The trajectories were analyzed with MDAnalysis\cite{2011_MDAnalysis}. Structural figures were all generated using VMD\cite{1996_VMD}.

\subsection{Experiments}

\paragraph{Reagents} Erastin (S7242) and Fer1 (S7243) were purchased from Selleck.Rabbit monoclonal anti-GPx4 (ab125066) and Rabbit monoclonal anti-$\upbeta$-actin (ab213262) were purchased from Abcam and Anti-rabbit IgG, HRP-linked Antibody (7074) were purchased from Cell Signaling Technology.

\paragraph{Cell lines} 786-O cells (Procell) were cultured in RPMI1640 medium supplemented with 10\% FBS. 786-O GPx4-knockout cells were grown in RPMI1640 supplemented with 10\% FBS and \qty{1}{\micro M} ferrostatin-1.

\paragraph{Protein expression and purification} ALOX15, GPx4 and GPx4 mutants were expressed using Escherichia coli expression system as previously described\cite{2016_ALOX15,2019_GPx4_experiment}. In brief, ALOX15 with a His-tag was purified using a nickel-nitrilotriacetic acid column (HisTrap HP, GE Healthcare) and an anion-exchange column (Q Trap HP, GE Healthcare). GPx4 and mutants were fused with His-tag and purified using a nickel-nitrilotriacetic acid column at 4 °C. The eluted enzyme was further subjected to dialysis for 12 h [100 mM Tris-HCl (pH 7.4), 1 mM DTT, and 20\% (v/v) glycerol] to remove the bulk of the imidazole. The final purity of the enzyme was >95\% as confirmed by sodium dodecyl sulfate–polyacrylamide gel electrophoresis. Protein concentrations were measured using Nanodrop 2000 (Thermo Scientific).

\paragraph{Mutagenesis experiments} All mutagenesis experiments were performed according to the instructions of the Mut Express II Fast Mutagenesis Kit (Vazyme). The pQE-30 bacterial expression plasmid and pcDNA3.1 mammalian expression plasmid of the U46C mutant of human cytosolic GPx4 was mutated to obtain the mutants. The DNA sequences of all mutants were verified by DNA sequencing. The protein expression and activity assays of the mutants were performed as described for the U46C mutant.

\paragraph{Preparation of liposomes and liposome leakage assay} Liposomes were prepared as previously described \cite{2021_MembraneDamage}. 16:0 PC (DPPC), and 18:0-20:4 PE (SAPE) were obtained from Cayman Chemical. The phospholipids (80\% DPPC and 20\% SAPE) were dissolved in chloroform as a 10 mg/ml stock solution. The phospholipid films were obtained by mixing a 100 $\upmu$L stock solution and 500 $\upmu$L of chloroform in round-bottomed 10 mL flasks and evaporating the solvent using a vacuum rotary evaporator. The lipid films were hydrated with 500 $\upmu$L of buffer L (50 mM PBS, pH 7.3, 100 mM NaCl) and the liposomes were subjected to 20 rounds of extrusion through a 100 nm polycarbonate membrane using a mini-extruder (Avanti). To prepare the 5-Carboxyfluorescein encapsulated liposomes, the lipid films were hydrated with 500 $\upmu$L of buffer CL (50 mM PBS, pH 7.3, 100 mM NaCl, 20 mM 5-Carboxyfluorescein). After performing the extrusion, the liposomes were purified by a desalting column to remove any external 5-Carboxyfluorescein, after which they were re-suspended in 500 $\upmu$L of buffer L. All the liposomes were stored at 4 °C and used within 24 h.

\paragraph{Liposome leakage assay} The reactivity of ALOX15 with phospholipids in liposomes has been reported \cite{1996_ALOX15}. For the liposome leakage assay, 10 $\upmu$L of the liposomes were diluted into 100 $\upmu$L of buffer LA [100 mM Tris-HCl (pH 7.4), 3 mM glutathione, 100 mM NaCl]. When 50 $\upmu$M of ALOX15 and 50 $\upmu$M of GPx4 were added to the reaction mixture, the fluorescence signal was measured by a microplate reader. The fluorescence signal was then recorded at 10-s intervals and the relative activity of different mutants was compared by calculating the change of fluorescence signal intensity within ten minutes.

\paragraph{GPx4 activity assay} By coupling the oxidation of NADPH to NADP$^{+}$ by oxidized glutathione in the presence of glutathione reductase, we assessed GPx4 activity by measuring the decrease in NADPH fluorescence emission at 460 nm (excitation at 335 nm). Purified GPx4 mutants (50 $\upmu$M) were preincubated in the assay buffer A [100 mM Tris-HCl (pH 7.4), 5 mM EDTA, 0.1\% (v/v) Triton X-100, 0.2 mM NADPH, 3 mM glutathione, and 1 unit of glutathione reductase] for 5 min at 37 °C. Then the reaction was initiated by adding tert-butyl hydroperoxide (25 $\upmu$M). Fluorescence signals were recorded for 6 min on a plate reader (Synergy, BioTek). 

\paragraph{CRISPR–Cas9-mediated gene knockout} 786-O GPx4-knockout cells were generated by CRISPR–Cas9 technology using transfection reagent  (Lipofectamine 3000, Thermo Fisher) with PX458. The GFP-positive cells were sorted using a BD FACSAriaIII Fusion flow cytometer after two days and seeded in 96-well plates (1 cell per well). The single clones were grown in 96-well plates for at least 3 weeks before detecting the protein expression levels. To target GPx4, the guide sequences preceding the protospacer adjacent motif were: guide 1, 5'-CAACAACAAGTCCGCACGTC-3'; guide 2, 5'-TGTTCCACGCGCGCGGGTCG-3'.

\paragraph{Overexpression of GPx4} Codon-optimized human GPx4 gene with a GPx4 U46C mutation was synthesized (GENEWIZ) and cloned in the expression vector pcDNA3.1.GPx4 U46C/L130A, GPx4 U46C/L130S and U46C/L130Q were subsequently generated using site-directed mutagenesis.

\paragraph{Cell viability assay} The cells were seeded in 96-well plates at 3,000 cells per well in triplicates and allowed to adhere overnight. On the next day, cells were treated with erastin. Cell viability was assessed 48 h after the treatment using MTT (Sigma-Aldrich). MTT (0.5 mg/mL) was added to the cells and the plates were incubated at 37 °C for 4 h. Then, the growth medium was removed and DMSO was added to dissolve the precipitate. Absorbance was measured at 570 nm using a plate reader (Synergy, BioTek).

\paragraph{Quantification and statistical analysis} Statistical information for individual experiments can be found in the corresponding figure legends. Values were presented as mean ± s.d. P values for pairwise comparisons were calculated using the two-tailed t-test. Quantitation and statistics were calculated GraphPad Prism 9.

\paragraph{Western blotting} Cells were washed twice with PBS, lysed in RIPA-lysis buffer, combined with 1 $\times$ SDS loading buffer, and incubated for 15 min at 98°C. Protein extracts separated on 10\% gels and transferred onto PVDF membranes. Membranes were incubated for 24 h in PBS with 0.1\% Tween-20 (PBST) containing 5\% bovine serum albumin (BSA) (Sigma Aldrich) and primary antibodies. After washing with PBST, membranes were incubated at room temperature for 30 min in 5\% BSA and PBST containing secondary antibodies.

\section{Acknowledgements}

This work was supported by the Science Fund for Creative Research Groups of the National Natural Science Foundation of China (T2321001). Part of the MD simulations were performed on the Computing Platform of the Center for Life Sciences at Peking University.










\clearpage
\onehalfspacing

\begin{table}[htbp]
\centering
\caption
    {\textbf{Information of simulated systems.} Four types of phospholipid bilayers were constructed for our simulations: a pure POPC membrane, a mixed POPC/POPE membrane, a mixed POPC/POPE/POPS membrane, and a peroxidized PC/PE/PS membrane. The composition of the peroxidized PC/PE/PS membrane is POPC 10\%, POPE 20\%, POPS 20\%, HPPC 10\%, HPPE 20\%, and HPPS 20\%.}
\begin{tabular}{>{\centering\arraybackslash}m{1.25cm}|>{\centering\arraybackslash}m{2cm}>{\centering\arraybackslash}m{2cm}>{\centering\arraybackslash}m{2cm}>
{\centering\arraybackslash}m{1.5cm}>{\centering\arraybackslash}m{1.75cm}>{\centering\arraybackslash}m{2.5cm}}
\toprule
\textbf{System ID} & \textbf{Protein} & \textbf{Membrane} & \textbf{Bead numbers} & \textbf{Box size (nm$^\mathbf{3}$)} & \textbf{Simulation time (μs)} & \textbf{Replicate numbers} \\ \midrule
\textbf{CG-1} &GPx4-U46C &POPC 100\% &18264 &$12\times12\times15$ &50 &1\\ \midrule
\textbf{CG-2} &GPx4-U46C &POPC 50\% POPE 50\% &18292 &$12\times12\times15$ &50 &1\\ \midrule
\textbf{CG-3} &GPx4-U46C &POPC 20\% POPE 40\% POPS 40\%&18175 &$12\times12\times15$ &50 &1\\ \midrule
\textbf{CG-4} &GPx4-U46C &POPC 100\% &18175 &$11.92\times11.92\times14.66$ &10 &3 (mutate PE and PS to PC based on 50-μs result of CG-3)\\ \midrule
\textbf{CG-5} &GPx4-U46C &peroxidized PC/PE/PS membrane&18693 &$12\times12\times15$ &50 &1\\ \midrule
\textbf{CG-6} &GPx4-U46C &peroxidized PC/PE/PS membrane&18693 &$12\times12\times15$ &10 &6$\times$6 (six different initial configurations)
\\ \midrule
\textbf{AA-1} &GPx4-U46C &peroxidized PC/PE/PS membrane&211625 &$12.27\times12.27\times14.13$ &0.5 &3 (L130 inserted in the membrane)\\ \midrule
\textbf{AA-2} &GPx4-U46C &peroxidized PC/PE/PS membrane&211621 &$12.28\times12.28\times14.14$ &0.5 &3 (L130 was close to the membrane surface)\\ \midrule
\textbf{AA-3} &GPx4-U46C/L130S &peroxidized PC/PE/PS membrane&222617 &$12.27\times12.27\times14.13$ &0.5 &3 (S130 inserted in the membrane)\\ \midrule
\textbf{CG-7} &GPx4-U46C &peroxidized PC/PE/PS membrane&18693 &$12.27\times12.27\times14.13$ &10 &3\\ \midrule
\textbf{CG-8} &GPx4-U46C/L130A &peroxidized PC/PE/PS membrane&18692 &$12.27\times12.27\times14.13$ &10 &3\\ \midrule
\textbf{CG-9} &GPx4-U46C/L130S &peroxidized PC/PE/PS membrane&18693 &$12.27\times12.27\times14.13$ &10 &3\\ \midrule
\textbf{CG-10} &GPx4-U46C/L130Q &peroxidized PC/PE/PS membrane&18693 &$12.27\times12.27\times14.13$ &10 &3\\
\bottomrule

\end{tabular}
\label{tab:DifferentMembrane}
\end{table}

\clearpage
\onehalfspacing

    \begin{suppfigure}[htbp]
        \centering
            \includegraphics[width=12cm]{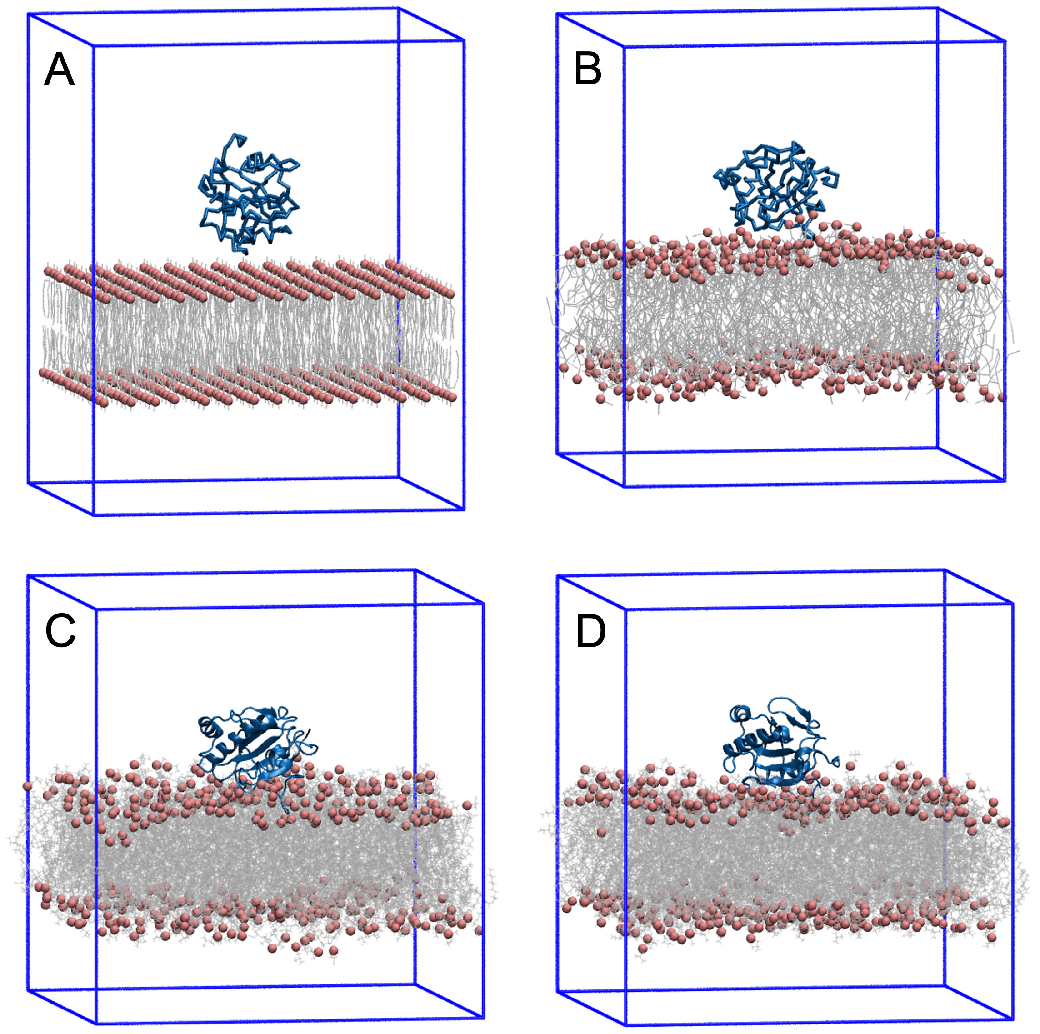}
             \caption{The simulation systems. (A) The initial system and (B) the final system of CG MD simulations. (C) The initial system and (D) the final system of AA MD simulations. Explicit water molecules and ions are omitted from these figures for clarity; detailed system compositions (including protein types and lipid compositions) are provided in Table S1.}
            \label{fig:MD_systems}
    \end{suppfigure}

\clearpage
\onehalfspacing

    \begin{suppfigure}[htbp]
        \centering
            \includegraphics[width=15cm]{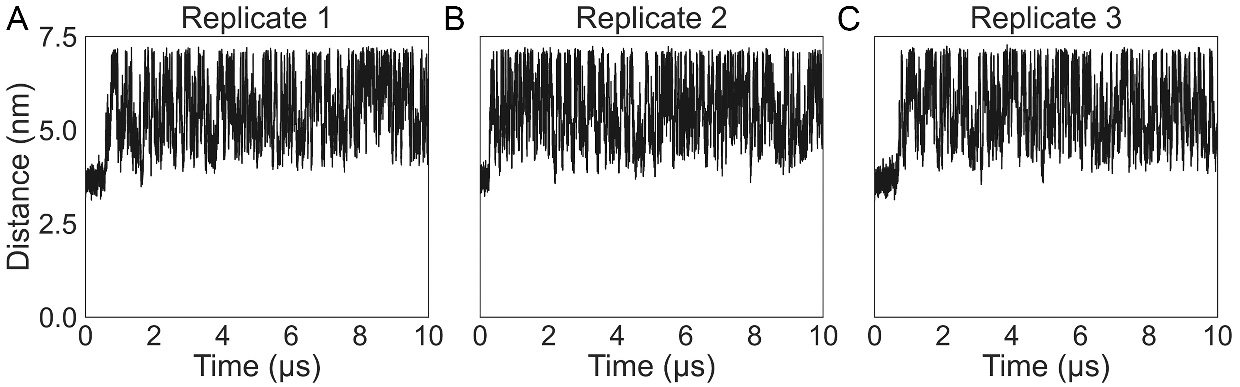}
             \caption{Distances between GPx4 and the POPC membrane in the extended CG MD simulation (three replicates). The initial configuration was based on the stable binding pose of GPx4 on the POPC/POPE/POPS membrane after the 50-μs CG simulation, with the head groups of PE and PS converted to PC (System-ID: CG-4 in Table S1). The distance is calculated between the center of mass (COM) of the protein and the membrane midplane and is always fluctuating and mostly greater than 4 nm after 1 μs, indicating that GPx4 was detached from the mutated membrane.}
            \label{fig:PCPEPStoPC}
    \end{suppfigure}

\clearpage
\onehalfspacing

    \begin{suppfigure}[htbp]
        \centering
            \includegraphics[width=15cm]{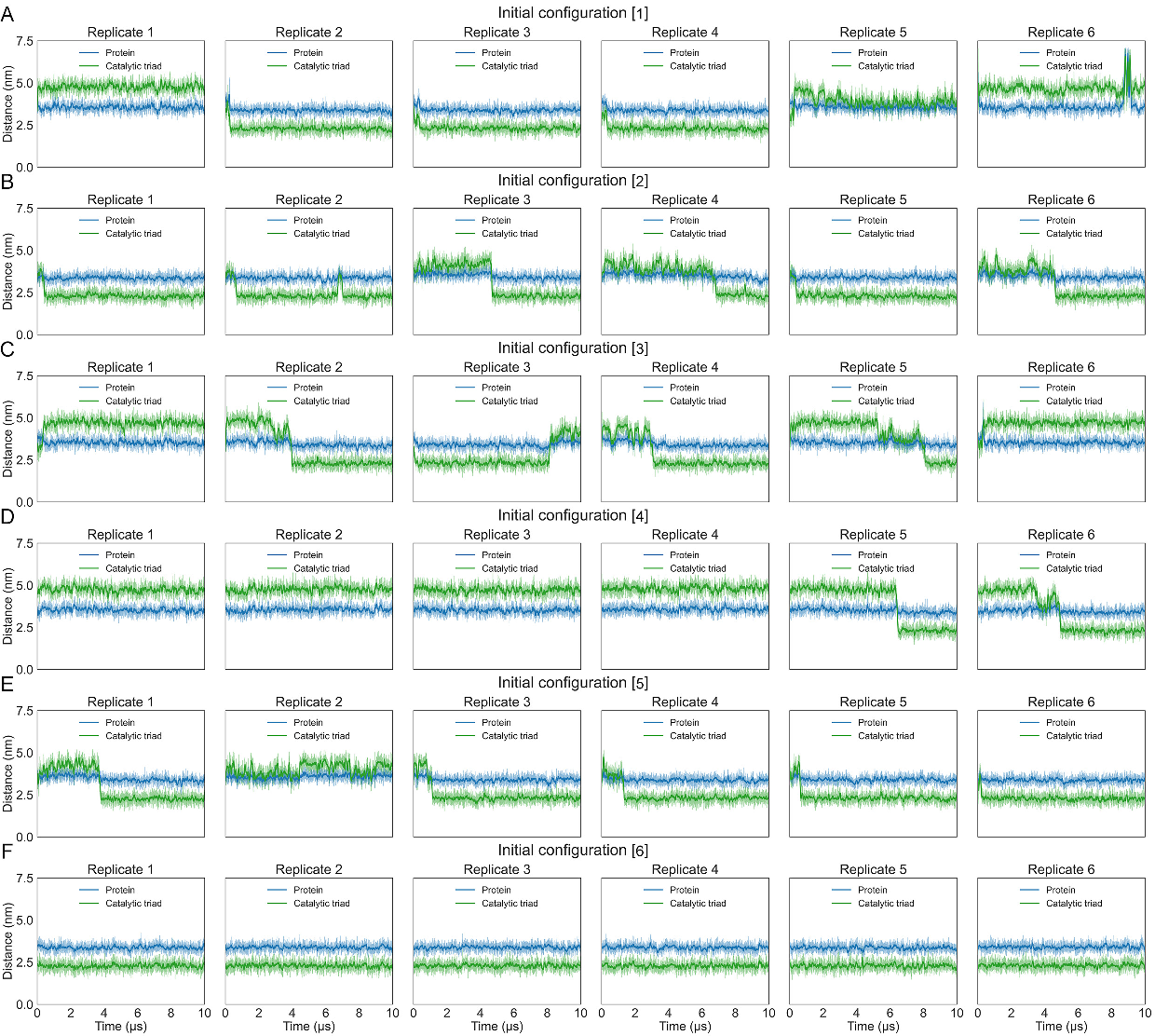}
             \caption{Evolution of binding between GPx4 and membranes with various starting configurations. The distances between the membrane midplane and the center of mass of GPx4 or the catalytic triad were calculated to indicate the binding process. Six replicate simulations were conducted for each of the six initial configurations (indicated by the cubes in Fig. 3A, System-ID: CG-6 in Table S1). 25 out of these 36 trajectories converged to the same stable binding interface within the simulation time. In this interface, the catalytic triad is in direct contact with the membrane surface (green line located at around 2.5 nm from the membrane midplane).}
            \label{fig:Rotations}
    \end{suppfigure}

\clearpage
\onehalfspacing

     \begin{suppfigure}[htbp]
         \centering
             \includegraphics[width=10cm]{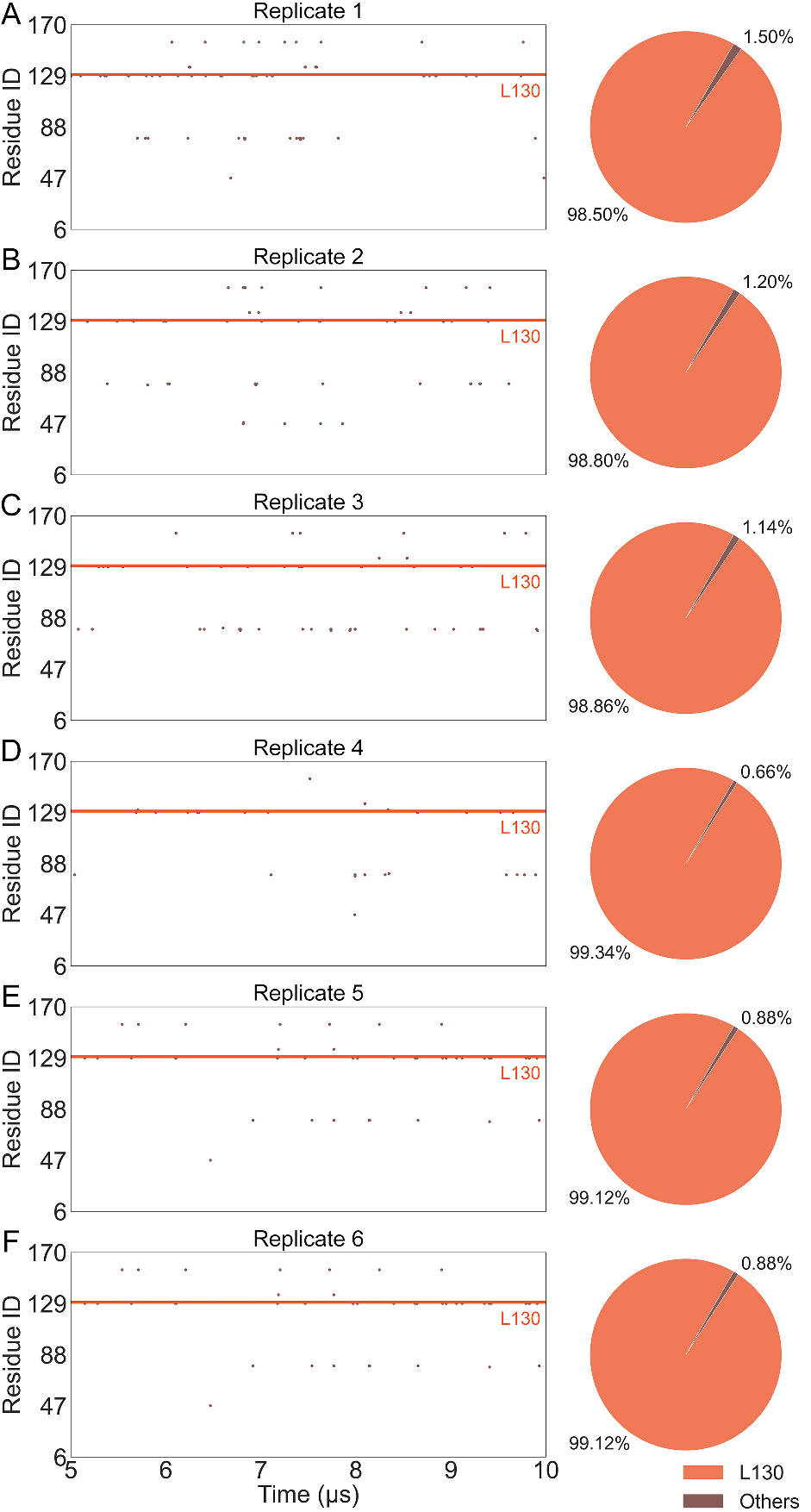}
              \caption{The residues closest to the midplane of the peroxidized membrane in six CG simulations (Fig. S3F) with the sixth initial configurations (Fig. 3A [6]). The closest residue in each frame is shown with a dot. The dot is shown in orange when the closet residue is L130. Otherwise, it is brown. As can be seen, the L130 was the closest residue to the membrane midplane in this stable binding mode.}
             \label{fig:min_ResID}
     \end{suppfigure}

\clearpage
\onehalfspacing

     \begin{suppfigure}[htbp]
         \centering
             \includegraphics[width=15cm]{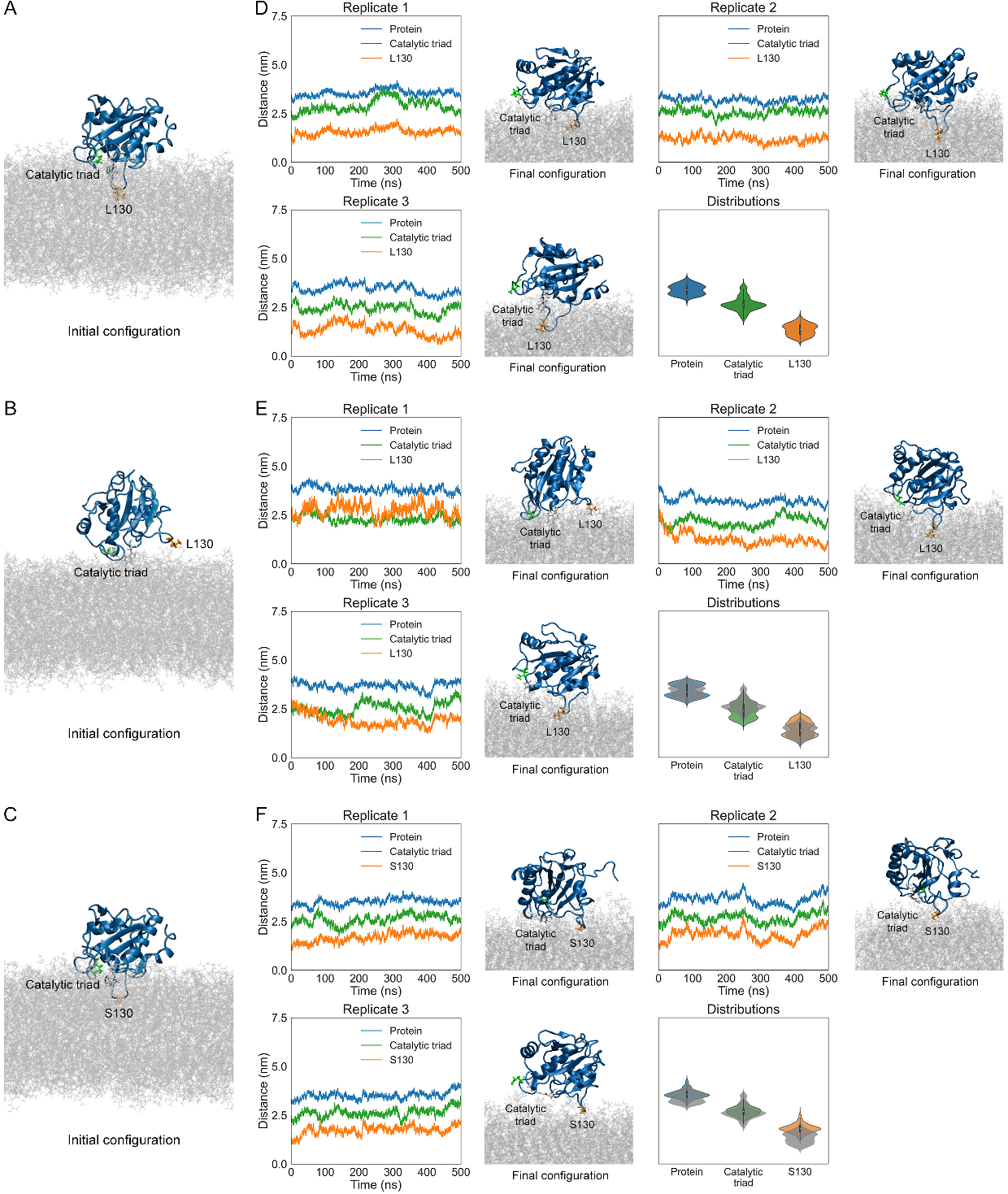}
              \caption{(A-C) Different initial configurations for all-atom simulations, with L130 inserting deeply into the interior of the membrane (A, System-ID: AA-1 in Table S1), L130 being close to the membrane surface (B, System-ID: AA-2 in Table S1), and S130 inserting deeply into the interior of the membrane in configuration (C, System-ID: AA-3 in Table S1). (D-F) Distances between the centers of mass of the protein, the catalytic triad, L130/S130 and the midplane of the peroxidized membrane in atomistic simulations with different initial configurations (A-C, three replicates for each configuration), as well as final configurations of each simulation and distributions of the distances. In panels (D) and (F), all three replicates are used to plot the non-grey distribution. However, since L130 fails to insert into the interior of the membrane in the first replicate of panel (E), only the second and third replicates are used to plot the non-grey distribution. The grey distributions shown in panels (E) and (F) are exactly the same as those presented in panel (D), which are used as references for comparison. These distribution of the distances are analyzed over the time range from 200 ns to 500 ns of each simulation replicate.}
             \label{fig:AA_simulations}
     \end{suppfigure}

\clearpage
\onehalfspacing

    \begin{suppfigure}[htbp]
        \centering
            \includegraphics[width=12cm]{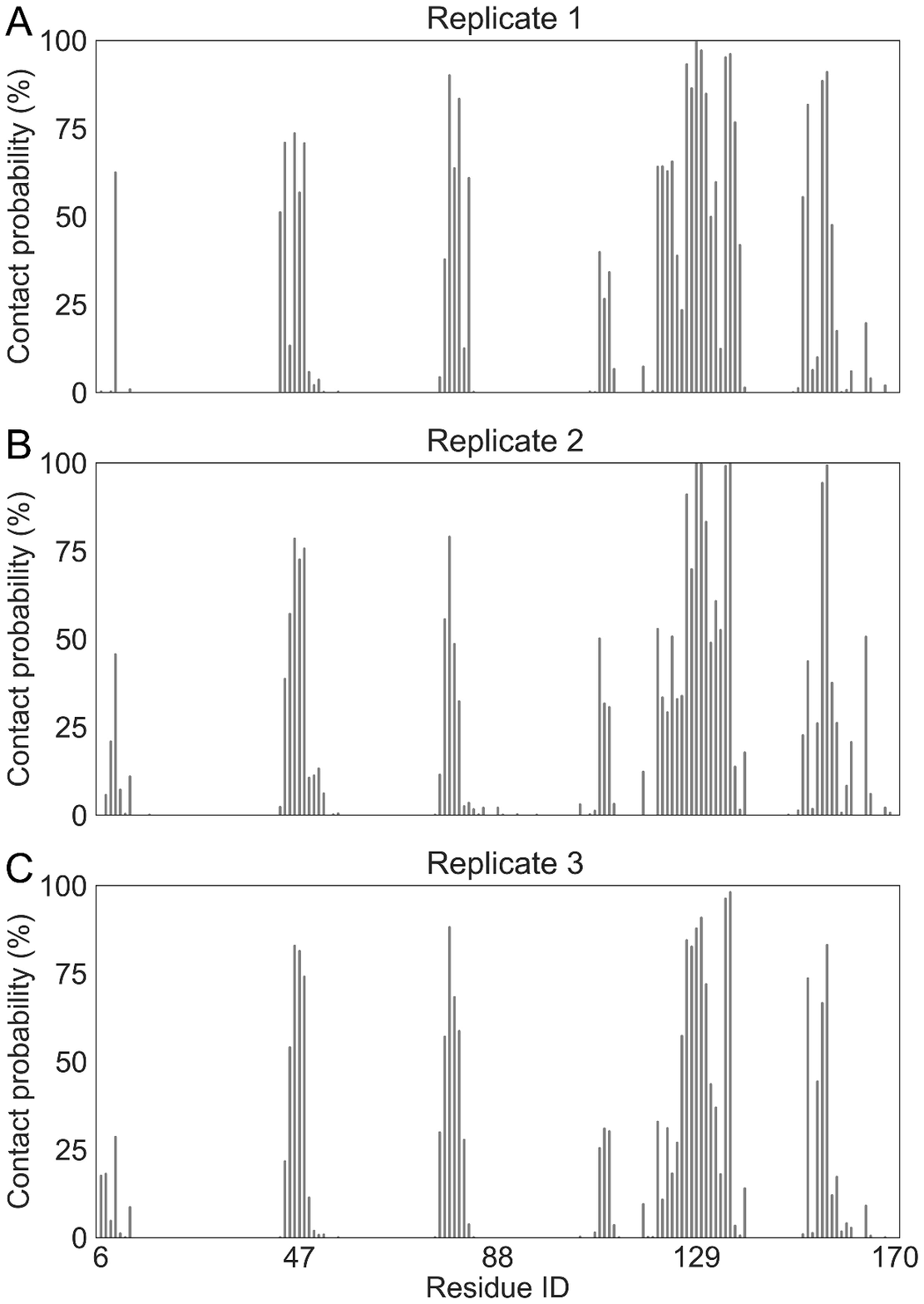}
             \caption{Contact probabilities of each residue with the tail chains of phospholipids in AA simulations (three replicates, System-ID: AA-1 in Table S1). The "contact" was defined as the distance between any heavy atom of the residue and any heavy atom of the tail chains being less than 6 Å. The contact probability of L130 was 97.08\%, 99.60\%, and 90.82\% in the three replicate simulations, respectively.}
            \label{fig:Residue_contact}
    \end{suppfigure}

\clearpage
\onehalfspacing

    \begin{suppfigure}[htbp]
        \centering
            \includegraphics[width=15cm]{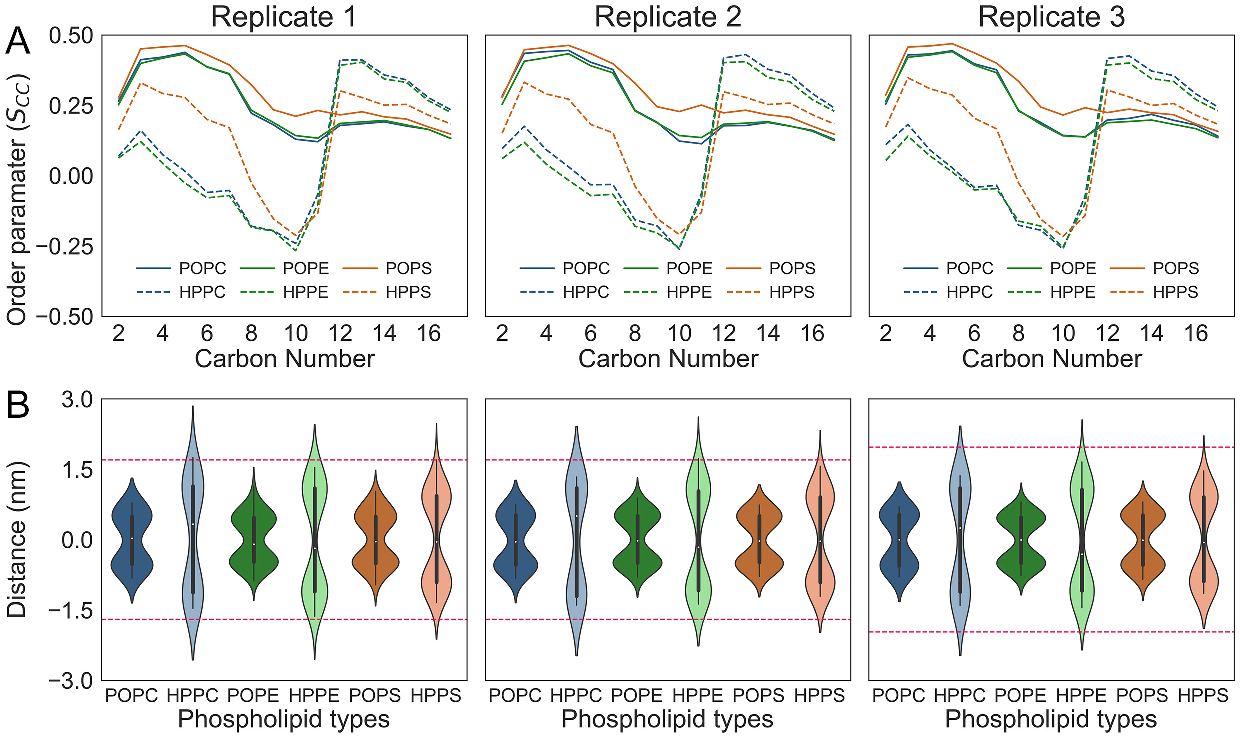}
             \caption{(A) Order parameters of various lipids in the AA simulations (three replicates, System-ID: AA-1 in Table S1). The order parameters are calculated for the peroxidized chains of HPPC, HPPE, and HPPS, as well as for the corresponding chains of POPC, POPE, and POPS. For the calculation of the order parameter of the \(i^{\text{th}}\) carbon atom, it is determined by using the angle between the line connecting the \((i - 1)^{\text{th}}\) and \((i + 1)^{\text{th}}\) carbon atoms and the $z$-axis. As can be seen, the peroxidized chains of peroxidized phospholipids are significantly more disordered than the corresponding chains of non-peroxidized phospholipids, especially the part between the head group and the peroxidized group (carbon number <11). This part tends to be perpendicular to the $z$-axis, which can lead to exposure of the peroxidized group. (B) Distributions of the distances between the carbon atoms connected to peroxidized groups of peroxidized phospholids or the corresponding carbon atoms of normal phospholids and the midplane of the peroxidized membranes in AA simulations (three replicates, System-ID: AA-1 in Table S1). The dotted line in the figure represents the average distance of the head groups of lipids ($\sim$1.97 nm). It is clear that the distributions of the carbon atoms connected to the peroxidized groups are closer to the surface of the membranes. The results support that the peroxidized group will be more exposed to the membrane surface due to the conformational change of the peroxidized phospholipids.}
            \label{fig:lipids_chains}
    \end{suppfigure}

\clearpage
\onehalfspacing

\begin{suppfigure}[htbp]
        \centering
            \includegraphics[width=8cm]{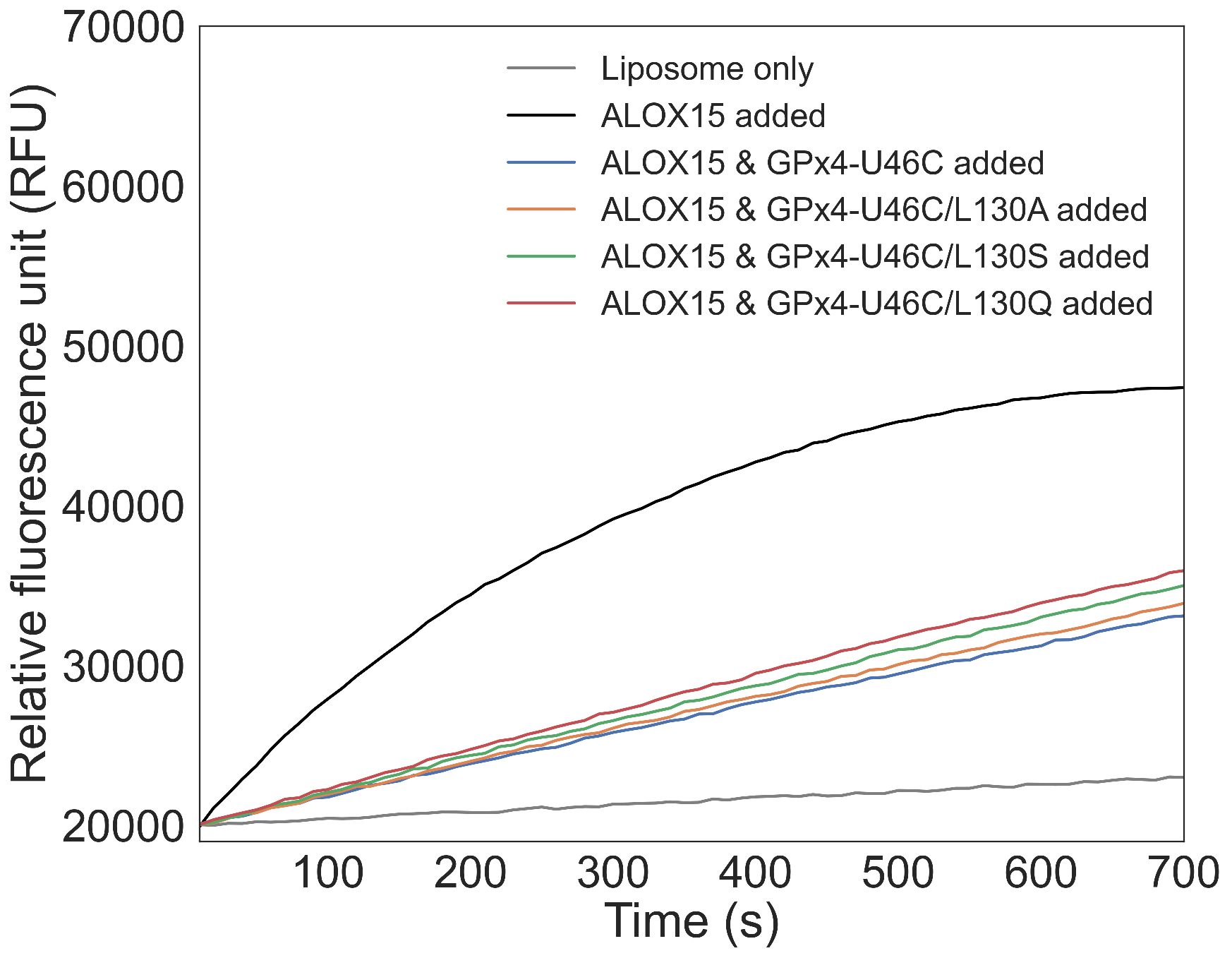}
             \caption{Time courses of liposome leakage caused by ALOX15. As shown, GPx4 and its mutants can block ALOX15-induced liposome leakage at different rates.}
            \label{fig:GPx4_experiments}
    \end{suppfigure}

\clearpage
\onehalfspacing

\begin{suppfigure}[htbp]
        \centering
            \includegraphics[width=12cm]{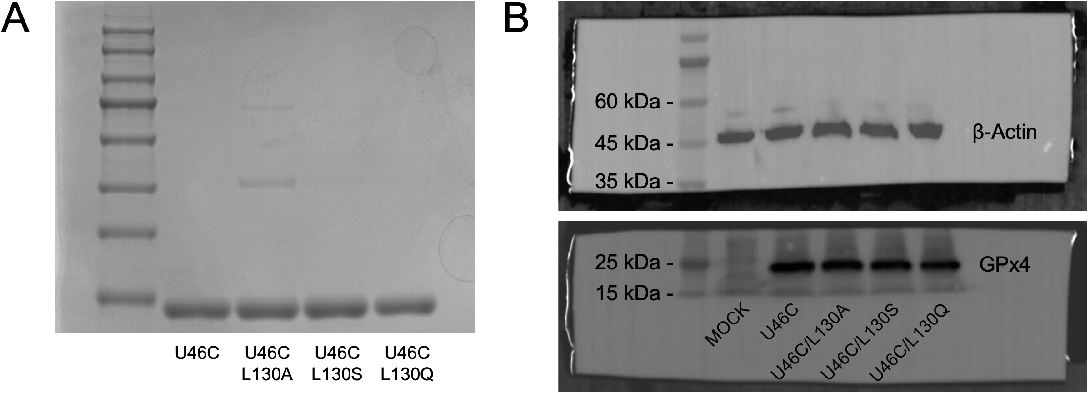}
             \caption{Characterization of GPx4 and its mutants. (A) SDS-PAGE of GPx4 and its mutants, with protein bands visualized by Coomassie brilliant blue staining. (B) Cellular protein expression levels of GPx4 and its mutants, determined by western blotting. The expression levels of GPx4 and its mutants were shown to be comparable.}
            \label{fig:GPx4_experiments_characterization}
    \end{suppfigure}

\clearpage
\onehalfspacing

\begin{suppfigure}[htbp]
        \centering
            \includegraphics[width=8cm]{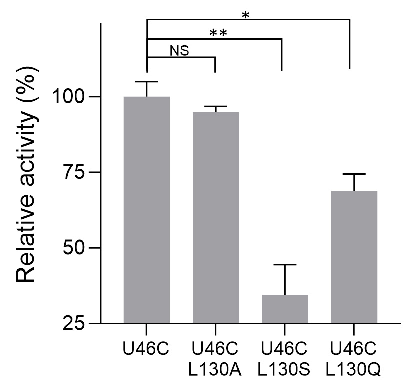}
             \caption{Relative activity of GPx4 and its mutants to inhibit ferroptosis normalized by quantifying the GPx4 protein levels determined by western blotting (Fig. 5B). The activity of U46C/L130A was similar to U46C, while U46C/L130S and U46C/L130Q were significantly inactivated.}
             \label{fig:GPx4_experiments_relative_activity}
    \end{suppfigure}

\clearpage
\onehalfspacing

\begin{suppfigure}[htbp]
        \centering
            \includegraphics[width=16cm]{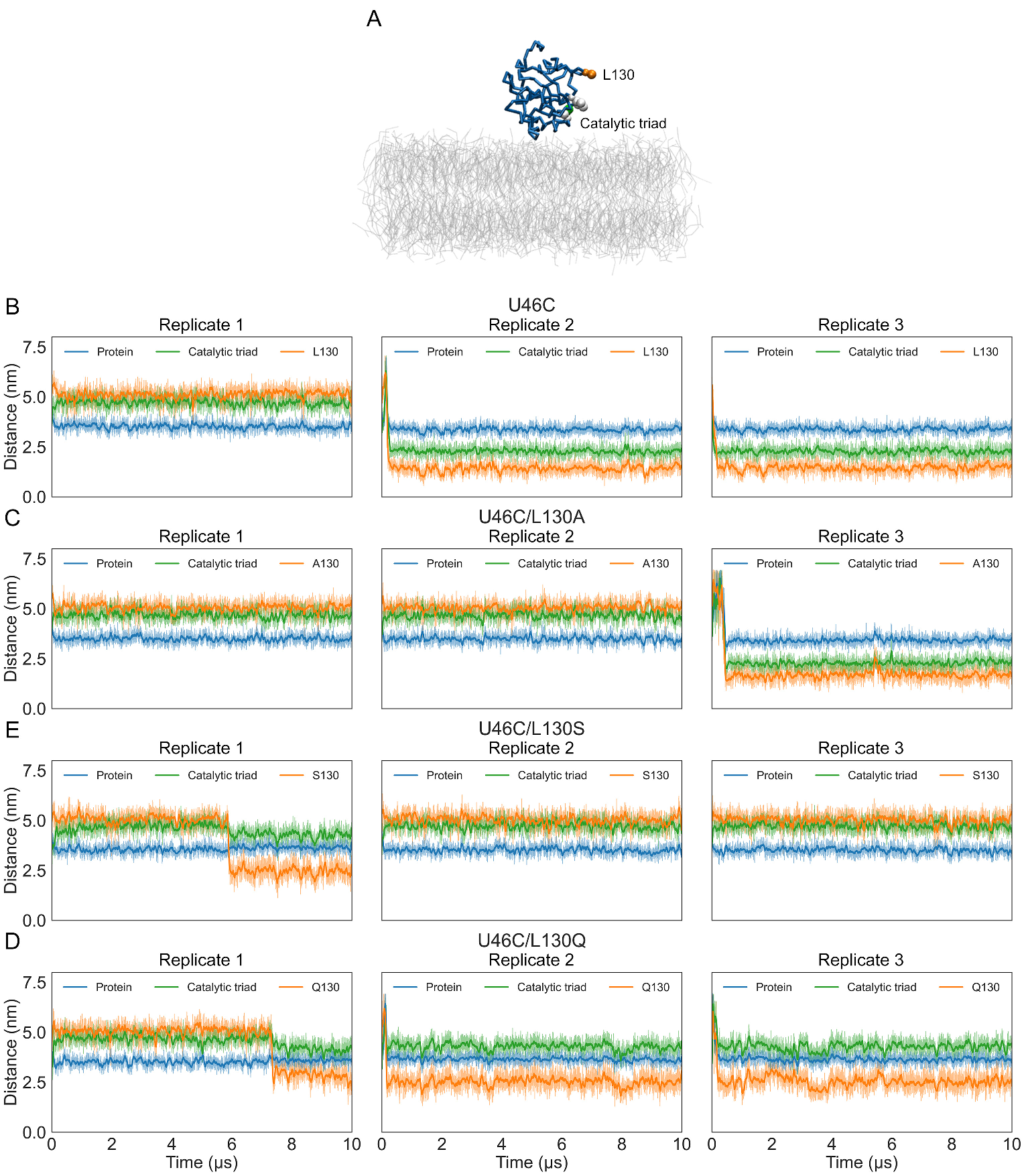}
             \caption{Binding of GPx4-U46C and its mutants on the membrane. (A) Initial configuration of GPx4-U46C with the peroxidized membrane (POPC:POPE:POPS:HPPC:HPPE:HPPS=1:2:2:1:2:2), where GPx4 was randomly placed in a position close to the membrane surface (System-ID: CG-7 in Table S1). For other mutants (U46C/L130A, U46C/L130S, and U46C/L130Q), the initial positions of all atoms were the same, except that the L130 residue was mutated into A130 (System-ID: CG-8), S130 (System-ID: CG-9), and Q130 (System-ID: CG-10), respectively.
             (B-E) The distances between the membrane midplane and U46C (B), U46C/L130A (C), U46C/L130S (D), and U46C/L130Q (E), as well as their catalytic triads and the 130th residues (three replicates for each mutants). The distance was calculated between the center of mass (COM) of the protein atoms and the membrane midplane.
             Only U46C (two out of three) and U46C/L130A (one out of three) were observed to bind to the peroxidized membrane and form an effective binding interface within the simulation time, in which the catalic triad is in direct contact with the membrane surface (green line at around 2.5 nm).}
            \label{fig:GPx4_mutant}
    \end{suppfigure}

\clearpage
\onehalfspacing
\bibliography{references}  

\end{document}